# Generation of Infra sound to replicate a wind turbine[1]

**Richard Mann, Associate Professor, Cheriton School of Computer Science, University of Waterloo, Ontario, Canada.**

**William Mann, Ancaster, Ontario, Canada.**

## Abstract

We have successfully produced infrasound, as a duplicate of that produced by Industrial Wind Turbines.

We have been able to produce this Infrasound inside a research chamber, capable of accommodating a human test subject.

It is our vision that this project will permit others, with appropriate medical training and ethical oversight, to research human thresholds and the effects of this infrasound on humans.

Our role has focused on producing the tools, systems, and hardware required, to permit this research to go forward.

This paper describes the evolution of our project from the original vision, through the construction of proof of concept prototypes, a series of improved models and their associated accessories /operating systems, to the final test chamber as it stands now ready to deploy.

Also included are the mathematical and computational data supporting our claim that infrasound conditions inside the chamber can be made to duplicate those from actual Industrial wind turbines at approved setback distances.

## Terminology

**Sound waves** are variations of air pressure relative to the local barometric pressure. Average barometric pressure is 101.3 kPa (kilo Pascals). 1 Pascal of variation corresponds to 94 decibels.

**Pascal** is a unit of measure for barometric pressure. We think of our normal atmosphere as approximately 100,000 Pascals.

---

[1] Document history: Document created and submitted to University of Waterloo, Feb 25, 2018. Revised and publicly released, Mar 2, 2018. Published as Tech report 2018-01 in Computer Science on Mar 3, 2018. Document modified Mar 27, 2018. Submitted to Biorxiv (Physiology) April 5, 2018. Rejected April 6, 2018. Submitted to arXiv May 2, 2018.

**Infrasound** refers to low frequency sound (pressure changes in air), that are below the range of human hearing. Human hearing is typically defined as from 20Hz to 20,000 Hz (Hertz, also known as cycles per second).

**Infrasound waves** are the same variations of air pressure relative to atmospheric but occur at much lower frequency levels (from 0.1hz to 20 Hz) and are below the audible level.

**Audible sound** is in the range of 20 Hz to 20,000 Hz.  Hz refers to the number of cycles per second

**Decibel** is the unit of measurement for sound. Decibels are a logarithmic scale. Adding decibels has the effect of multiplication of pressure levels (Pascals).  For example:

Adding 20 db corresponds to multiplication of pressure level (Pa) by 10
Adding 40 db corresponds to multiplication of pressure level (Pa) by 100
Adding 60 db corresponds to multiplication of pressure level (Pa) by 1000

## Traditional methods of producing infrasound in a chamber/room

**Traditional methods have involved either a sealed chamber or a modulated input.**

**Sealed chamber method**

The simplest method to generate infra sound is to use a sealed chamber.
Infra sound is generated by sealed speaker cabinets(s), placed inside the chamber.
As the speaker cone moves outward, the total occupied volume (speaker enclosure + diaphragm) increases slightly.
As the speaker cone moves inward, the total volume decreases.
Since the volume of the chamber is fixed, the chamber pressure will decrease inversely with the increasing speaker volume.
We note that for a large chamber volume, multiple speakers would be required, since each speaker has only limited displacement (typically 5 to 10mm for commercially available sub woofers).
Furthermore, this type of chamber requires a perfect seal, which is difficult to achieve.
Finally the sealed chamber must have no ventilation, so human experiments must be limited and carefully monitored.

A device of this type for Wind turbine simulation was described by Walker and Celano (2015)

A related idea is for the speakers to occupy a part of the chamber surface, such as the wall or ceiling. A similar principle applies, as the movement of speakers increases or decreases the volume of the chamber.  This type of device was used by Nussbaum and Reinis (1985) and most recently by Cooper et al (2017).

Finally, a variant of this method was used to provide infra sound stimulus directly to the ear, via a tube and a sealed ear piece, while performing simultaneous fMRI brain scanning. (Weichenberger et al, 2017).  Note that in this case, only the ear was stimulated, so vestibular or whole body effects cannot be evaluated.

**Unsealed chamber method**

A second approach to generating infra sound in an area is to blow air into (or suck air out of) a chamber, room, or other enclosure.
As air blows into the chamber, the pressure increases and sucking air out reduces the pressure.
To generate infra sound air is repeatedly pumped in and out of the chamber (or room) at the desired frequency.
Since these devices pump air, the listening chamber need not be completely sealed.
As long as the air flow of the fan is sufficiently strong, it can overcome the leaks in the room/chamber, and very high levels of infra sound can be generated.
This allows use in a wide variety of spaces, including listening rooms, performance spaces, etc.

In this technology, a "rotary woofer" or "woofer fan", uses a fan with a variable blade pitch (Park et al, 2009).

In the forward position the blades generate positive flow, while in the backward position they generate negative flow.  A neutral position generates no flow.

The blades are moved by a "voice coil", an electromagnetic system, just like a regular loud speaker.  To generate low frequency sound the fans are tilted in and out at the desired frequency.
The woofer fan is placed in a sound isolated chamber, outside the listening room, or may be incorporated into a ventilation system.
Most recently, Kevin Dooley Inc. have produced a miniature version of this device, using a very high speed fan (10000 rpm) in a small (approx. 6" diameter) enclosure (Dooley, 2015).

## A third method, our current project and the subject of this paper

In October 2015 funds were approved by University of Waterloo Office of Research for the purchase of a woofer fan type device from Kevin Dooley Inc.

When we were unable to complete this purchase, we considered construction of our own woofer fan, but quickly decided against this, due to the complexity of construction.

We then began working on a new approach.

## Description of our proof of concept device.

Our system uses a separate constant air source and modulates the output.

We use a fan with a fixed air flow that pressurizes a chamber to some (small amount) above atmosphere.

The inside chamber pressure is modulated by an exit valve responding to specific time and frequency commands as further outlined.

This paper describes the technical aspects of reproducing infrasound in a chamber, the associated methods and equipment needed, and the evolution from a proof of concept prototype to a full size chamber capable of accommodating a human test subject.

We determined that actual human testing and data collection would be left to others with appropriate medical and ethical oversight. Our role would be to create the tools which would permit this research to go forward

In August of 2013 R. Mann began working with John Vanderkooy on a method of measuring and recording infrasound produced by Industrial Wind Turbines. By April 2015, they had developed a reliable method of measuring and recording infrasound produced by a single turbine.
They needed a way to isolate a single wind turbine from other turbines and from (random) wind noise. Their method used an optical telescope fitted with a photo detector to obtain reference blade passage periods, recording these together with the microphone infrasound signal.

Using signal processing they were able to isolate the infra sound from a single turbine. They have successfully measured infra sound from several different individual turbines in Ontario. The work was published at Wind Turbine Noise 2015 INCE/EUROPE, in Glasgow Scotland in April 2015.

Having accomplished reliable measuring our next step was to move forward with the following objectives:

Produce Infrasound in a consistent and dependable manner that is a result of specific input commands.

Produce Infrasound duplicating that produced by industrial wind turbines.

Produce infrasound in a chamber capable of accommodating a human subject in a lab setting.

Produce infrasound which can be used for data analysis, data recording, and a study of environment.

## Evolution and description of our system from proof of concept device to present:

### Proof of concept System description:

The key components of the system are: Air pressure source, Acoustic chamber, Modulation valve, Sensing device(s), and data Input and measurement recording system.

**Air pressure source:**
We needed a source of air pressure to some increment over atmospheric that was consistent, continuous, and controllable. For the proof of concept prototype we used a Fasco Industries (Model 702110702) draft Inducer designed for exhausting flue gases from furnaces.
We initially controlled volume and pressure by venting off excess air on the high pressure side. For subsequent prototypes we used a new model of draft inducer that is speed controllable.

**Acoustic chamber**: We constructed a composite wooden box measuring 27 x 30 x 70 cm (approx. inside dimensions) giving 56.7 liters of volume storage. It was made to a degree of robustness such that changes of pressure within the scope of our experiment had no deformation effect. The acoustic chamber was sealed at all edges and corners to prevent any pressure leakage so that pressure changes within it were caused only as a result of our input.

**Modulation Valve:** We fitted onto the Acoustic chamber a modulation valve which is under command of our data input control. This valve can release chamber air to the outside environment based on volume and time interval commands thereby altering the pressure inside the chamber.

We constructed this valve by adapting a speaker which already has the characteristic of being able to move its cone in and out in response to electrical input in the process of making normal sound, and is also sensitive to commands below the level of human hearing in the infrasound range of 0 to 20 Hz.

By sacrificing speakers we were able to cast an exact mirror of the actual speaker cone which we call the "anvil".

After much experimenting we were able to mount this anvil in perfect concentric relation to a new duplicate of the ruined speakers, such that commands to the speaker cause the cone to move towards, or away from the anvil in a controlled manner. In the molding process we included an opening at the center of the anvil which became the point of air out flow.

The result is that the speaker became a valve restricting or increasing air flow between the cone and anvil in response to specific time and volume commands. We mounted this assembly to the acoustic chamber such that air pressure could be allowed to escape into the outside atmosphere under the control of our data input system. This had the effect of modulating the pressure inside the chamber in response to our input commands.

**Sensing device**: The MKS Type 223 Baratron measured the pressure difference between the inside of the acoustic chamber and the outside. This device (model 00001AABS-SPCAL) has a sensitivity of approximately 50 Pa /V. This was determined using a G.R.A.S. 40AZ infra sound microphone as a reference.

**Data Input and measurement recording system:** This system is PC based, using a National Instruments myDAQ data acquisition board.

In early May of 2016 we were able to report to Office of Research that we had completed the proof of concept prototype, and had successfully produced infrasound in the chamber.

We also reported data on input response and included bode plots documenting resulting conditions in the chamber.

Included in the May 2016 report was our original drawing describing the principle of the modulation valve, and a photo of the proof of concept device as shown below.

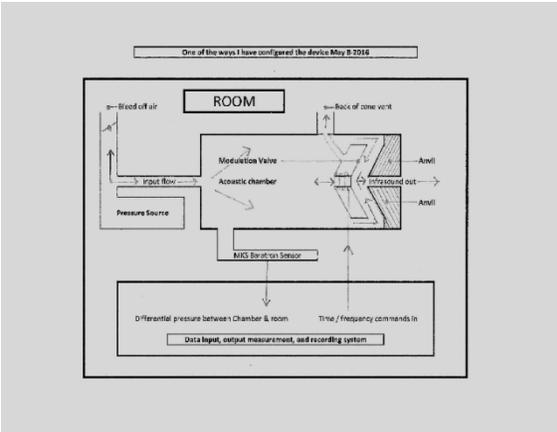 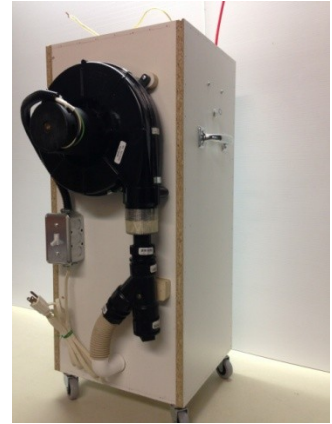

## Post proof of concept progress

Having determined that we could produce a computer controlled infrasound signal in a small proof of concept chamber, we scaled up to a chamber of approximately 200 liters and subsequently a third chamber of approximately 400 liters while continuing to refine our air source, modulation valve, and data input and recording methods. As well we continued to work on the mathematical and computational research necessary to achieve sound wave output to an exact duplicate of input data.

## Present state of the project as of this report

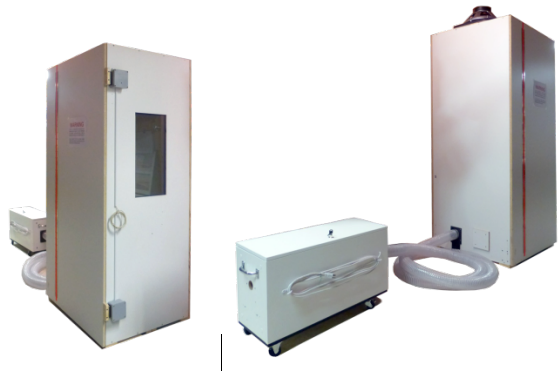

**Present chamber**:

We have now completed a full scale chamber system and related components. The latest chamber, capable of accommodating a human subject, has inside dimensions of 76cm X 87cm X 189cm resulting in an internal volume of approximately 1.25 cubic meters

The other system components (air pressure supply device and modulation valve) have also progressed in design based on the need to accommodate this new full scale. Designs have evolved incorporating the knowledge gained from previous prototypes, additional research, and trial and error.

**Present air pressure supply:**

After much experimenting with a variety of air pressure sources, we settled on a TORO centrifugal fan, model # 51619 designed to be operated at 115 Volts for use as a leaf blower.

Since we needed precise speed control for experimenting, we bypassed the supplied speed control, and regulated the fan output with a used Technipower brand Variac Autotransformer.

We found the optimum output for our purposes at a setting of 30 volts which provides a flow rate of approximately 25L/s at pressure of approximately 50Pa.

Some of the axial and cage fans we experimented with had unwanted performance changes due to feedback as a result of chamber modulation. The Toro centrifugal fan did not exhibit this tendency as it was designed for robust pressure as a leaf blower at 350 cfm (approx 165 L/s) and the fact that we are only using a very small percentage of its capacity for our needs.

The constant air source is enclosed in a cabinet separated from the modulation chamber by approximately 12 feet of 3 inch flexible hose normally used for workshop dust evacuation systems

**Current Modulation Valve**

Our current modulation valve consists of an Eminence brand model Eminator, a 1200 Watt 15 inch high power sub woofer, to which we fitted a cast anvil as described in the proof of concept description earlier in this paper.

**Safety features:**

Since human testing will be involved, safety was our number one priority, and the entire system design is based on the concept of "default to safe mode". The following are the safety design criteria incorporated to date.

Additional safety features will be added based on requirements of the end users of the system in accordance with their own health and safety requirements.

**Air pressure supply device:**

All mechanical components are contained in a locking cabinet to prevent unauthorized access.

The internal control devices are UL/CSA approved. A voltage controller (Variac) allows the air flow/pressure to be pre-set. The Air pressure system plugs into a standard grounded wall outlet, however, a relay allows the air supply to be turned on or off with standard HVAC 24 Volt wiring.

Since the pressure supply device and the chamber are two physically separate units the chamber and its associated controls have no connection to line voltage.

A second locking switch needs to be turned on by the operator before power is fed to the pressure supply device. Both keys are to remain in the possession of the system operator.

**The chamber:**

The chamber has no connection to line voltage.  All systems within the chamber operate on standard HVAC 24 Volts. There will also be a standard 12 Volt automotive outlet for laptop plug-in, if required.

An audio intercom system will allow for communication between subject and operator and a window enables visual contact between them.

The chamber entry door is designed mechanically to a default position of ajar. It is only held closed with electromagnetic latches and once power to the latches is interrupted the door is free to swing to the ajar position.

There is a prominently labeled STOP ring that the occupant can pull which will immediately release the door latches, allow the door to swing open, shut down the air pressure system, and immediately stop production of infrasound. There will also be a duplicate STOP ring on the outside for the operator's use in addition to the keyed master on / off switch.

A notice at chamber point of entry will prominently display a caution statement and indicates that the system must only be operated under direct supervision of appropriate staff. (Wording to be determined by end user's health and safety requirements.)

Sensors within the chamber (yet to be installed after discussion of requirements with the end user) will monitor air flow, air pressure and modulation levels. If sensors within the chamber detect any of the above to be outside of pre-determined limits, power to the door latches and air supply system will be immediately shut off.

Additional safety features will be added in accordance with the health and safety requirements of the final user.

## Performance data and technical description of the system

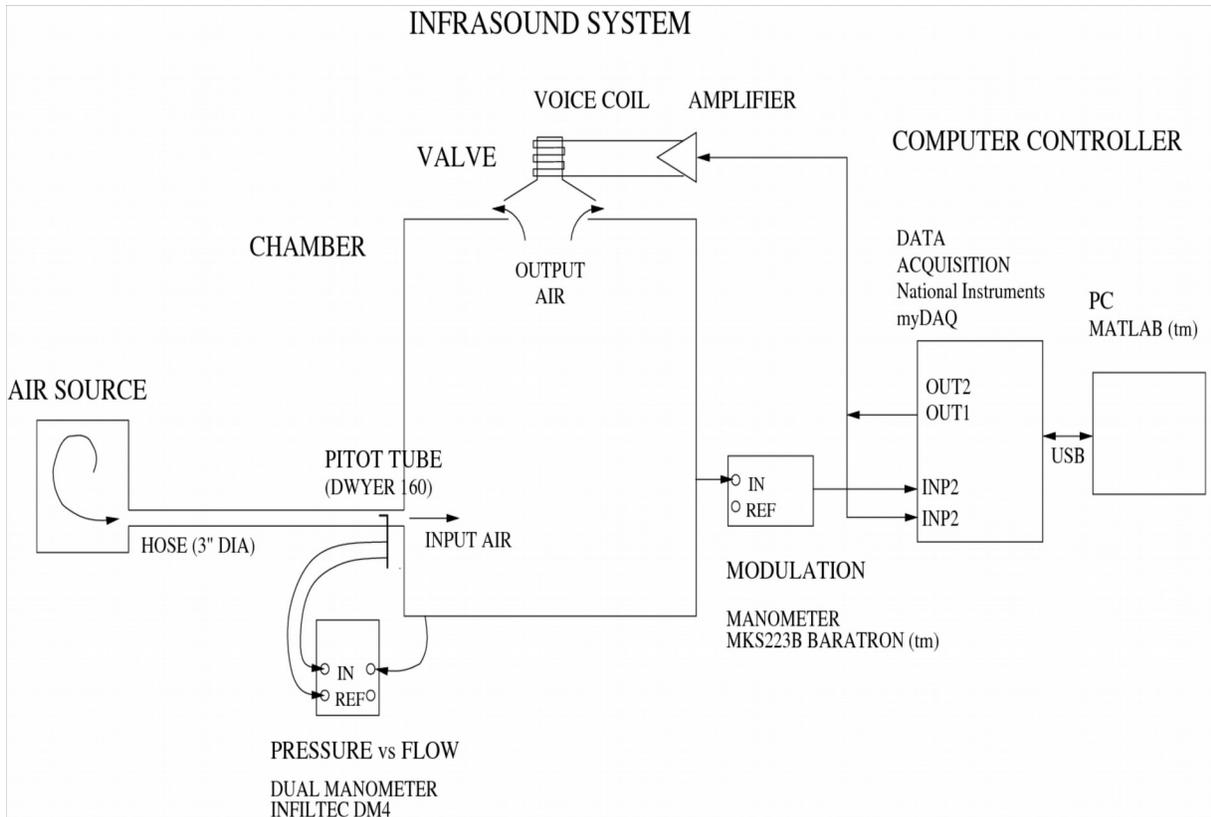

**Figure 1: System overview**

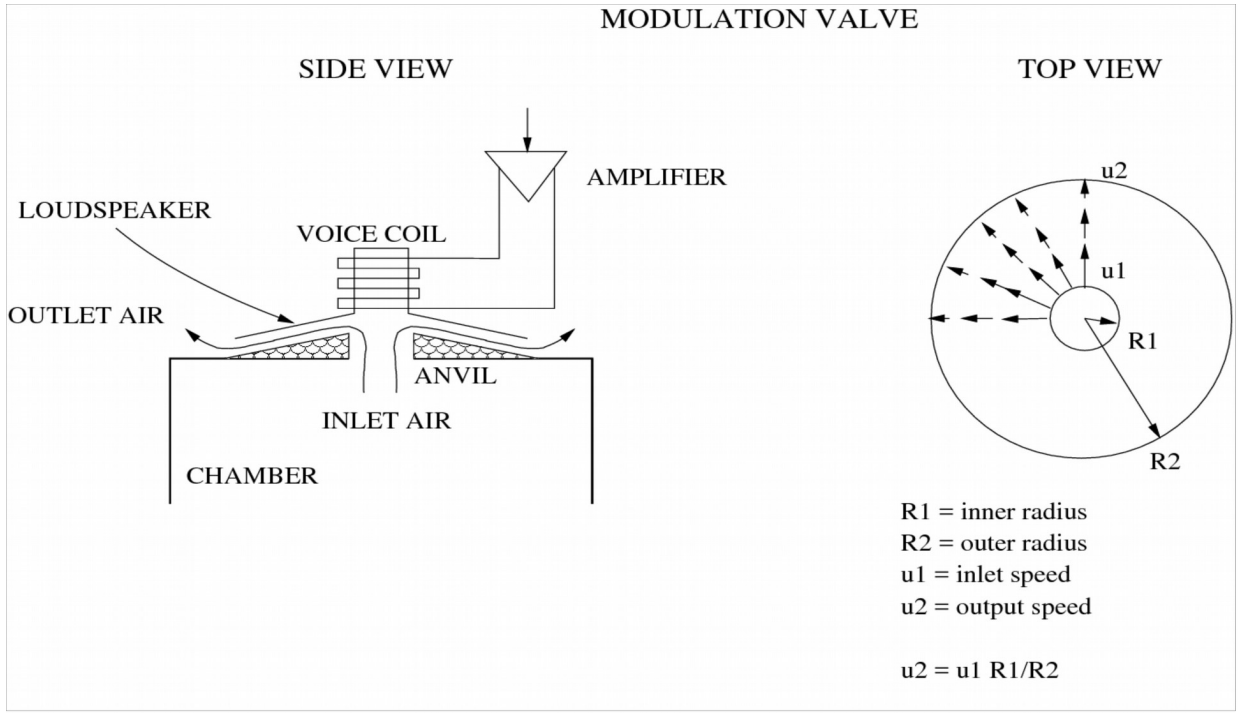

**Figure 2: Modulation valve**

## Properties of Pressure and Flow

Here we describe the properties of pressure and flow.
Pressure p, is the (over) pressure in the chamber, measured in Pascals (Pa).
Flow q is the air flow, measured in Liters per second (L/s).

The operation of the device is described by two curves, the "fan curve" and the "load curve" shown in Figure 3

These terms come from HVAC and ventilation systems where we want to have a fan or exhaust, and ensure there is sufficient pressure and/or air flow.

The "Fan Curve" shows the capabilities of the air source, in this case our blower.
At steady state, the system operates at a specific pressure and flow.
We refer to the "operating point" as a pair of numbers $(p, q)$.

When the valve is fully closed the air flow is zero and the pressure reaches a maximum.
This is shown as point $(p_{max}, 0)$ in the figure.

Alternately when the valve is fully open, we have a pressure of zero and a maximum flow of $q_{max}$. This is point $(0, q_{max})$ on the curve.

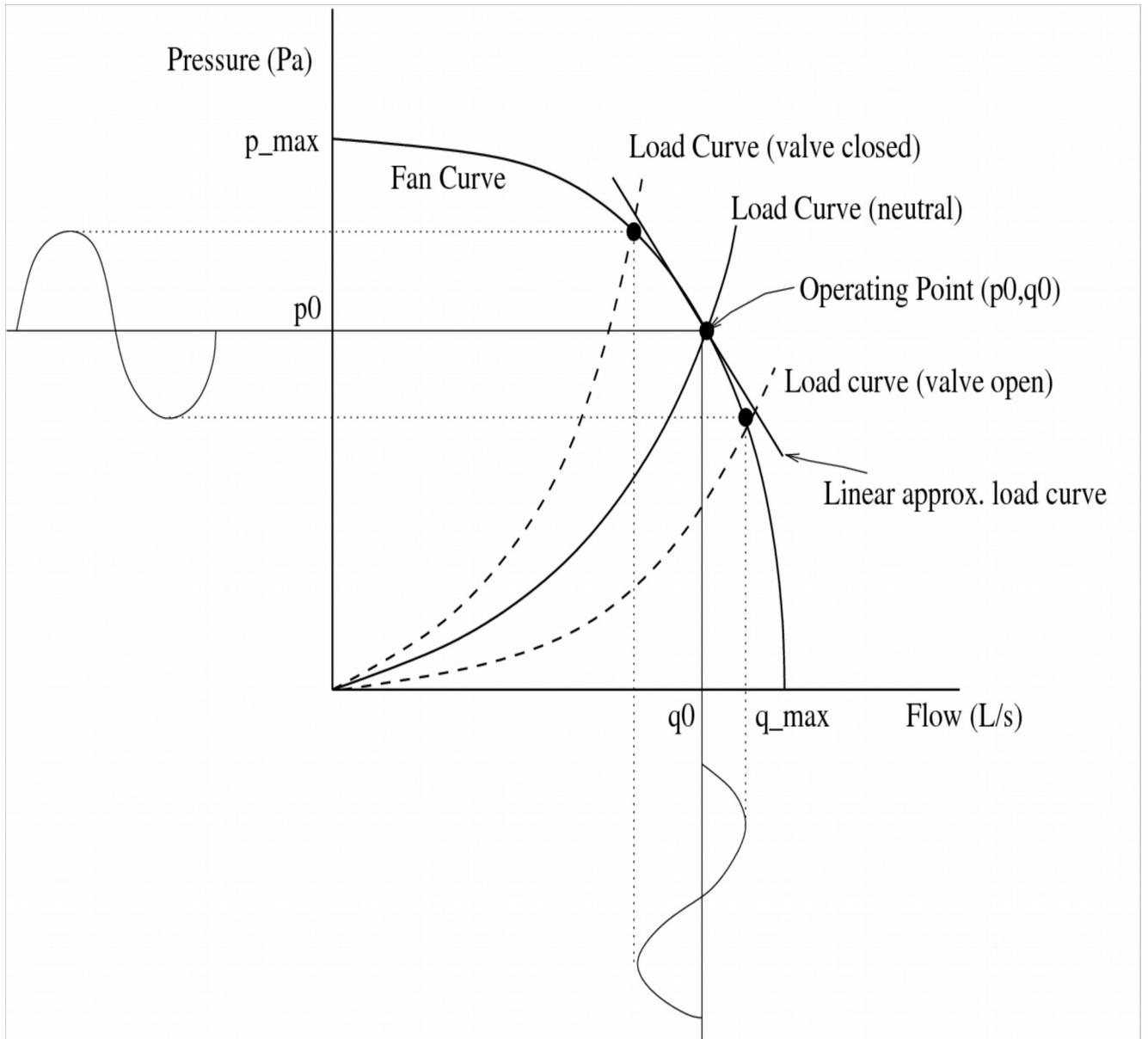

**Figure 3: System operating curve.** The system operates at the intersection of the fan curve and the load curve. The operating point depends on the load curve. The neutral position is denoted (p0,q0) in the figure. As the valve closes, pressure increases and flow decreases. As valve opens, pressure decreases and flow increases. The behavior is modeled with a local linear "load line". See text for details.

At any given operating point (p,q) we will be along the fan curve. The operating point is $(p_0, q_0)$.

As flow increases, pressure decreases, and vice versa.
Our aim is to choose an operating point, $(p_0, q_0)$ where there fan curve can be well approximated by a straight line.

In this way by linearly varying the flow (opening and closing the valve) we can generate a corresponding linear variation in the chamber pressure.

The second thing we need to consider is the load curve. The load curve is a quadratic function. Pressure on the vertical axis corresponds to the square of the flow on the horizontal axis. We have drawn here, the middle curve, the neutral curve where there is a moderate amount of flow. As we close the valve, the load curve moves upward, and as we open the value, the load curve moves downward.

If you combine these two things, the load curve and the fan curve, the actual pressure in the chamber will be at the intersection of these two curves.

We have shown here the neutral position with the valve half open, and have labeled the pressure and the flow as $(p_0, q_0)$. As we close the valve, we will move up the load curve and increase the pressure and as the valve opens we move downward and decrease the load pressure (the chamber pressure).

We can interpret the electro-acoustic system above as a purely electrical circuit using the following analogy:

$$(p - p_0) = \frac{\partial p}{\partial q}(p_0, q_0) \times (q - q_0)$$

$$V = R \times I$$

Where V is the voltage (pressure), I is the flow (current), and R is the acoustic resistance (at the load point).

In this way by linearly varying the flow (opening and closing the valve) we can generate a corresponding linear variation in the chamber pressure.

## Modulation valve

The main novel idea is the modulator which consists of a modified standard loud speaker as previously described in the proof of concept section.

The speaker presses its cone against a vent which we call the "anvil".
When the cone is fully pressing against the anvil venting is closed, and as the cone is moved away from the anvil venting increases.

Air from the chamber comes through the center hole of the anvil, passes between the anvil surface and the speaker cone and escapes to open air.

The top view of the modulator (figure 2) shows that the air begins exiting at the center of the valve at a high speed and as it comes to the periphery of the valve it slows down.
Finally when it exits the valve, it releases kinetic energy slowing down to zero.

The system works by Bernoulli's principle. That is, the (over) pressure in the chamber is equal to the kinetic energy of the air released from the valve. The formula is
$$p = 1/2 \rho u_2^2$$
Where $p$ is the pressure in Pa, $\rho$ is the density of air (kg/m^3) and $u_2$ is the speed (m/s) that the air exists the valve. Given an (approximately) fixed flow, as the valve closes the speed increases and the pressure increases. Likewise as the valve opens, the speed decreases, and the pressure decreases.

The speaker, (now an electro mechanical device called the modulator), controls the exit flow, and therefore the pressure in the chamber. Pressure in the chamber moves along the pressure curve in response to the input voltage on the valve.

## Pressure vs. flow – finding the operation point

To characterize the fan curve we measure the chamber pressure using a differential pressure sensor (MKS 223B Baratron) which measures the difference between the chamber pressure and the outside pressure. Flow (air speed) is measured by a Pitot tube in the 3" inlet tube.

The Pitot tube is a vent pointed towards the direction of flow that measures air pressure, (called dynamic pressure), and it records the difference between that, and the static pressure, which is perpendicular to the direction of flow.
The formula to determine airspeed is $u = \sqrt{(2/\rho) p}$ where $u$ is the air speed (m/s), $p$ is the pressure (Pa) and $\rho$ is the density of air (approx 1.225 kg/m^3 at 15 degrees C and at sea level).

Note that the flow is proportional to the square root of p as shown before for the load curve.

To compute the air flow we take the air speed in meters per second and multiply it by the cross sectional area of the inlet pipe (3" diameter).

To determine the fan curve we stimulate the system. We turn on the fan at the rated flow (30V in this case). We gradually move the speaker towards the anvil beginning at negative 5 volts of input to the speaker coil (full open) down to positive 5 volts (full closed) in increments of 0.5 volts.

As we do this we measure both the pressure and air flow at each point (Figure 4).

When the valve is fully open the pressure is very low and the flow is relatively high. As the valve gradually closes the pressure climbs and the flow declines.

In this example we noted incremental pressure increases from approx 0 to 175 Pascals and flow decreases from approx 27.5 Liters per second to approx 7.5 Liters per second as the valve incrementally closed.

Figure 5 shows the pressure/flow samples for all positions of the valve.  This is the "load curve".

So far we've have described the method by where we can increase and decrease pressure in our chamber by opening and closing our modulation valve,

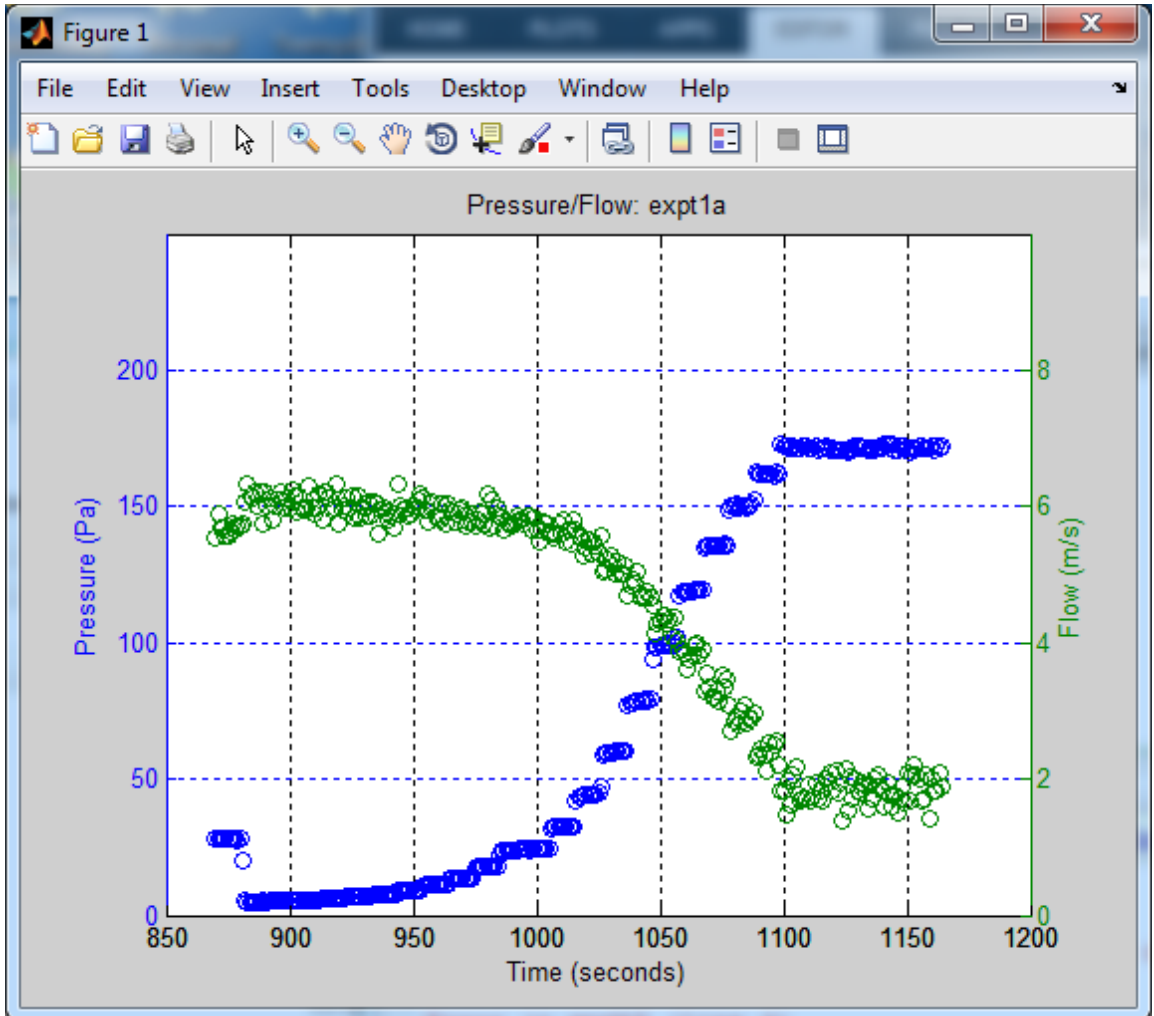

**Figure 4.  Measurement of fan response.  The modulator input is varied from -5.0 to 5.0V in 0.5V steps.  Each condition is held for ten seconds.  Pressure and flow measurements are shown as a function of time.**

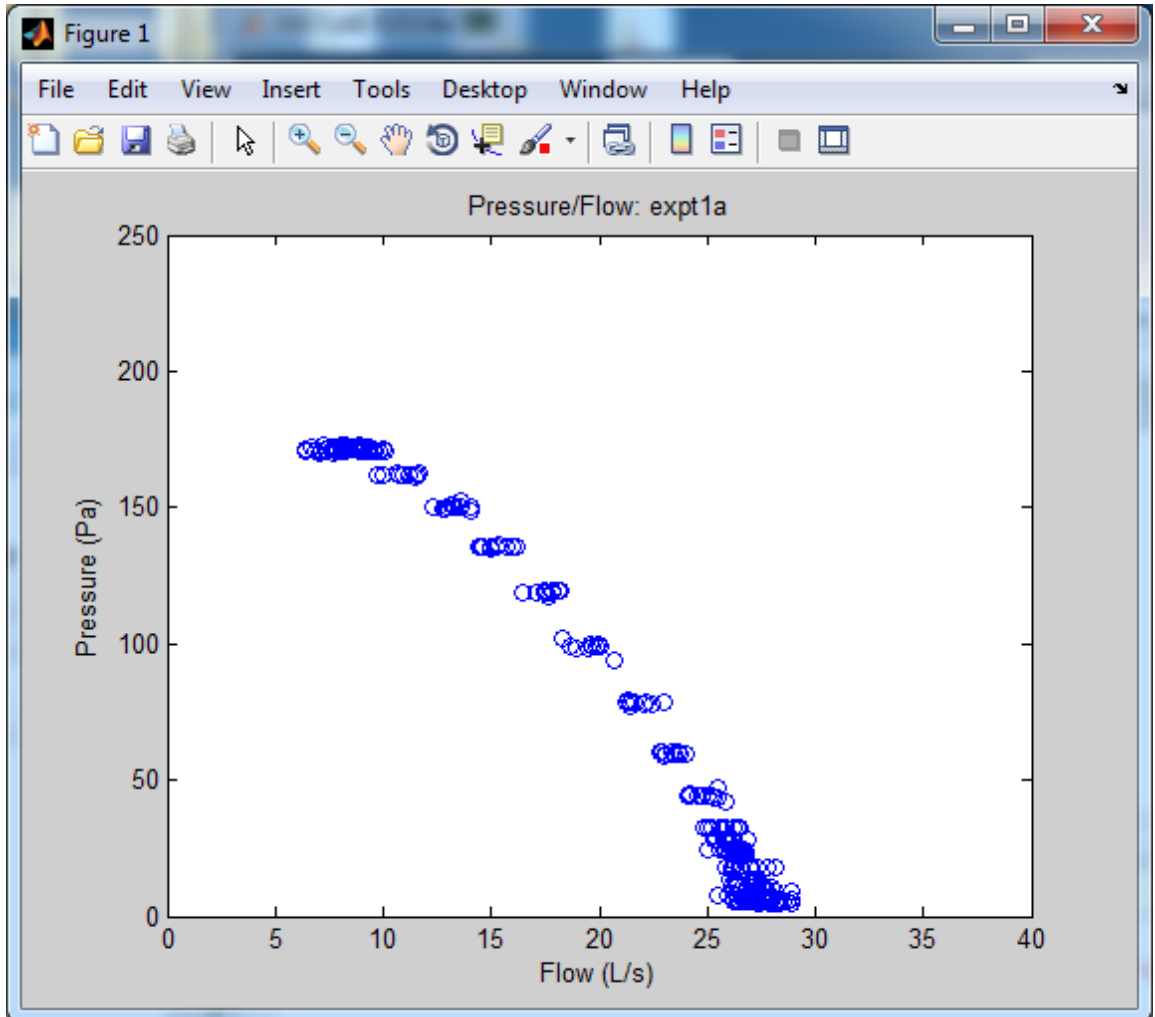

**Figure 5.** Pooling of pressure and flow measurements from Figure 4. This is the fan operating curve.

## Determining the System Frequency Response

The ultimate goal of this project was to generate an exact mirror of a real turbine pulse in a research chamber capable of holding a human subject.

This would allow others, with appropriate medical training and ethics approval, to research human exposure thresholds and reactions to this infrasound, while using our system as a research tool.

Anything less than a true duplicate of an actual turbine pulse within the chamber would compromise this research.

Since we needed to generate a pulse of arbitrary shape in the chamber, in addition to creating the appropriate output, we also needed to determine how our particular chamber would react as the frequency changed. Since a bigger chamber is slower to respond than a smaller chamber, and different chambers/rooms have differing characteristics, we needed to develop methods of measuring and compensating for, the frequency response of a given chamber/room.

In order to determine the system frequency response we will start by operating at approximately 50 Pascals over atmospheric pressure. Looking at the graph (figure 4) 50 Pascals pressure corresponds to an air flow of between 20 and 25 liters per second.

We note that if we open and close the modulating valve too quickly, the chamber has some delay in responding due to its own volume. If we vary the modulator very slowly, then the input pressure and air flow will follow the fan curve.

If we move the modulating valve in and out very quickly however, there will be almost no pressure change in the chamber, as there is not enough time to alter its relative amount of air.

Since a bigger chamber is slower to respond than a smaller chamber we need to measure the response of a given chamber.

Since our end goal is to generate a pulse of an arbitrary shape in the chamber, we must first determine how this particular chamber reacts as the frequency increases. This is called the chamber's frequency response.

A full study would analyze both the chamber and the air source separately, deriving a frequency response for each. For now we will treat the air source and chamber as a single unit and measure their combined frequency response.

The system is evaluated by measuring the response to pure tones (sine waves) of varying frequencies. At each test frequency a pure tone (sine wave) is sent to the modulator. For each frequency twenty (20) cycles are sent. In order to avoid transients (sharp changes in level), the beginning and the end of the signal are tapered with a cosine window of two cycles. Therefore, at each test frequency we have sixteen (16) complete unadulterated cycles to evaluate performance of the system.

Figure 6 shows the tone that's put into the modulator at a frequency of 0.8 hertz, which is the fundamental frequency of a wind turbine. 0.8 Hz means that there is 0.8 = 4/5 of a wave length within one second. If we turn that over it means that the period of this wave is 5/4 = 1.25 seconds. To test the chamber we put in 20 wave lengths at 1.25 seconds each which will take a total of 25 seconds. The blue curve shows this input into the modulator.

What we did in this case is put a starting value into the modulating valve of one volt creating a neutral position of the valve cone of 1 volt or 50 Pascals of allowed pressure in the chamber.

We then introduce a varying signal into the modulator that goes up to 1.5 volts and down to .5 volts having the effect of moving the cone slightly closer to the anvil, back through the neutral position, then slightly further away in amounts of 0.5 volts, repeating this cycle every 1.25 seconds.
This has the resulting effect of allowing air to escape from the chamber in response to the input commands to the modulation valve.
The red curve shown is the pressure measurement of the chamber and we see that as the modulator voltage goes up and down, the measured chamber pressure varies accordingly and in synchrony with that sign wave, and it has the same shape because we are in the linear region of the operating curve.

To measure the modulation, the manometer used is the MKS 223 Baratron and it has a sensitivity of 50 Pascals per volt. Since the red curve has a mean value of about one volt, that means there are about 50 Pascals in the chamber. The magnitude going up and down from the baseline is about 0.5 volts peak to peak, which corresponds to a deviation of 0.25 = 0.5/2 volts from the average value.

If you take 0.25 V and multiply by the sensitivity of the Baratron (50 Pa/V), you get 12.5 Pascals, so it is going up and down 12.5 Pascals in the chamber. This shows successful modulation in the chamber. In other words, a pure sine wave input to the modulator yields a pure sine wave variation in the pressure in the chamber.

We mentioned that there is a problem when we move the valve quickly. Figure 7 shows a pure tone at 8 Hz (cycles per second). In this case, what we observe is that 20 cycles of this wave now takes 1/10th of the time, or 2.5 seconds but still gives clean wave cycles.

Here we input the same wave that goes from one volt up to 1.5 and down to 0.5 volt so that's going to walk along the fan curve. Now look at what the chamber pressure is. The chamber pressure, (the static pressure), is about one volt or 50 Pascals, but the wave height has decreased because the chamber takes longer to respond. So in this case, we go up only by about 0.1 volt and down by about 0.1 volt.

We went from 0.25 volts in the previous example down to 0.1 volt so we went from 12.5 Pascals at 0.8 hertz to 5 Pascals at 8Hz.  Now also notice there is a delay inherent in this chamber.  There is a lag time.  As the valve opens, it takes a while for the pressure to respond and this is because of the volume of air in the chamber and that's called a 'phase delay'.
Now what we can do is characterize the performance of this chamber by taking measurements at several different frequencies.

What we are going to do is take measurements at 0.8 hertz and all multiples of that. So we are going to take a measurement at 0.8; 1.6; 2.4; 3.2; 4.0; and we are going to go all the way up to 25 times frequency and that corresponds to 20 Hertz.   At each step, we record 20 cycles and measure the magnitude and phase delay in each of those cases.  The results are shown in Figures 8 and 9.

We have already done two measurements. We have already done 0.8 hertz at the far left and we have already done 8.0 hertz here. This is shown on a logarithmic scale and we are just going to measure at each point.

Each dot is a measurement. The second harmonic is twice the first and then three times, four times, etc. now on the left is the gain in decibels and that is the relative gain between what we put into the modulator in volts and what we got out in volts from the pressure sensor.

Now, we mentioned the delay, and in a plot, we can also plot delay. This shows both the phase and degrees of delay so knowing a full wave length is 360 degrees, we know how much we shift, and we show a relatively smooth curve in the figure.

The plots we have shown here are called 'bode plots'.  By looking at the magnitude and phase response we can understand the system in terms of its mechanics or its electro acoustics. These can be measured.

We are not going to analyze this bode curve, we are only presenting it as an experimental result but now that we know this Bode curve, we can use this information to generate a turbine pulse, the subject of the next section.

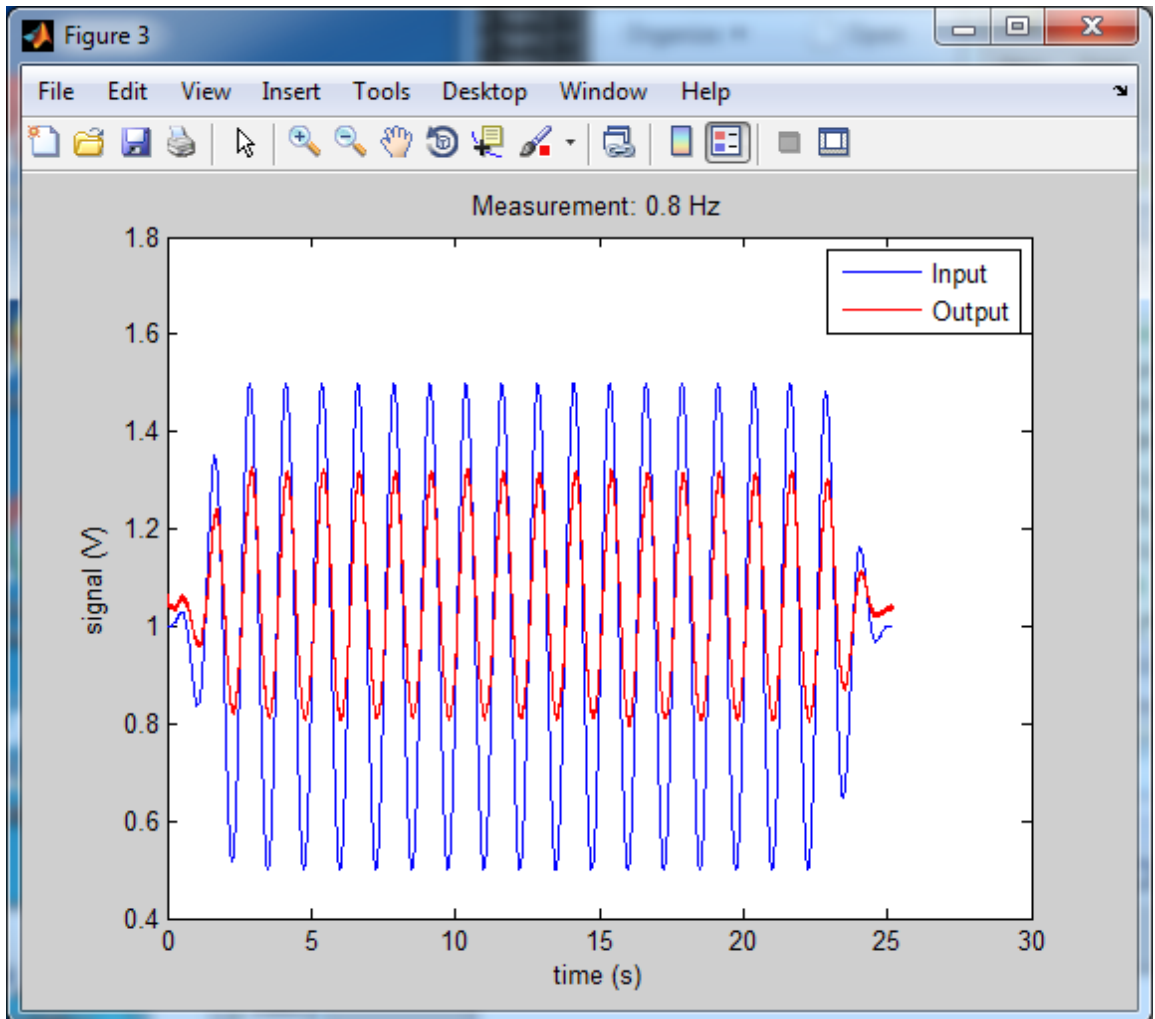

**Figure 6. Experiment results for wave of 0.8Hz. The input curve (blue) is the modulator input (in volts). The output curve (red) is pressure measurement in the chamber. The pressure is measured by a monometer (MKS 223B Baratron) with a sensitivity of 50 Pa/V. From this calibration value we see the chamber has an overpressure of approx. 50Pa = 1.0V). The output wave has a magnitude of approximately 0.25 V from the mean. This corresponds to approximately 0.25V * 50Pa/V = 12.5 Pa.**

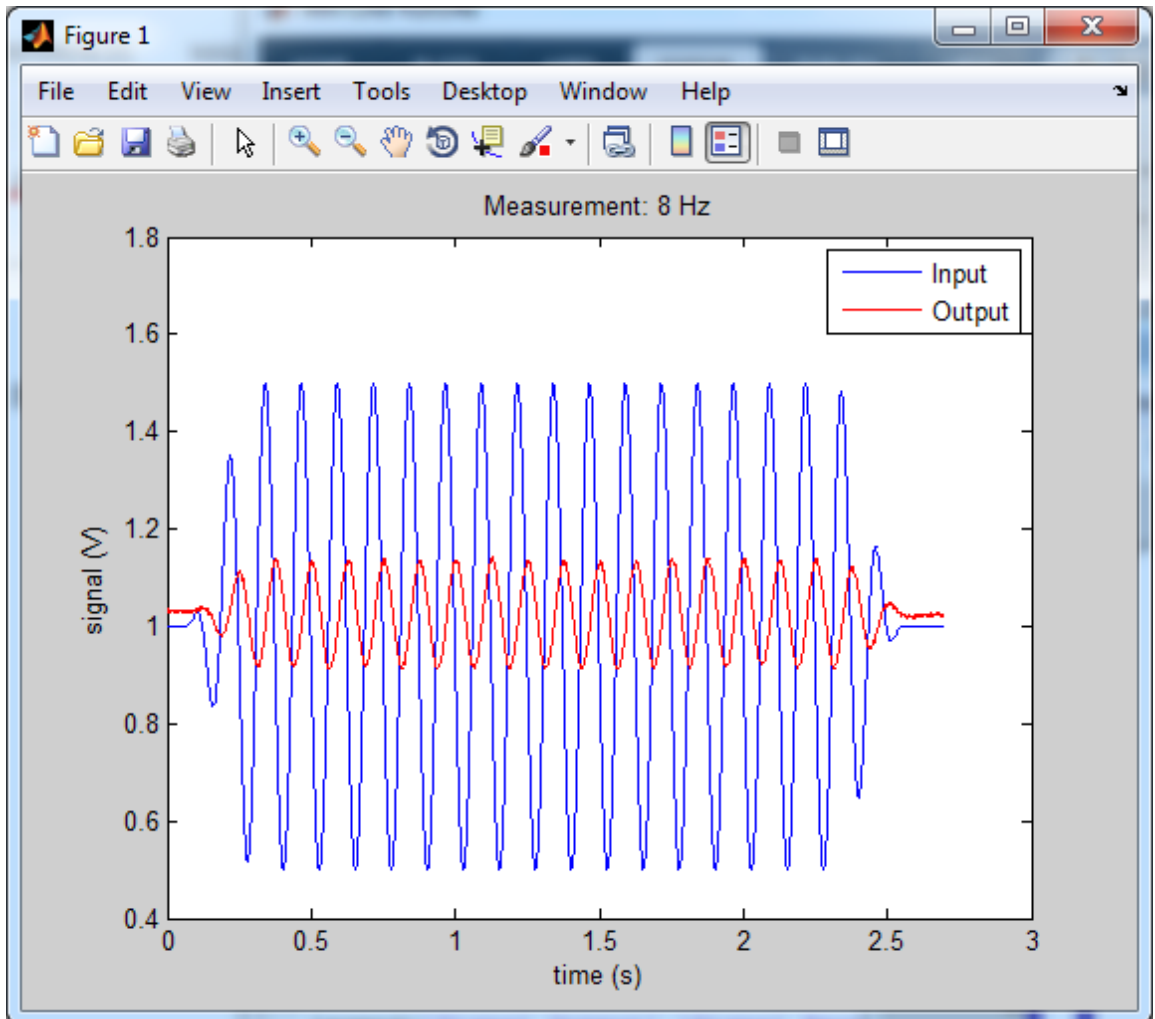

**Figure 7. Chamber response for 8Hz tone. The input curve (blue) is the modulator input (in volts). The output curve (red) is pressure measurement in the chamber. The pressure is measured by a monometer (MKS 223B Baratron) with a sensitivity of 50 Pa/V. From this calibration value we see the chamber has an overpressure of approx. 50Pa = 1.0V). The output wave has a magnitude of approximately 0.1 V from the mean. This corresponds to approximately 0.1 V * 50Pa/V = 5 Pa.**

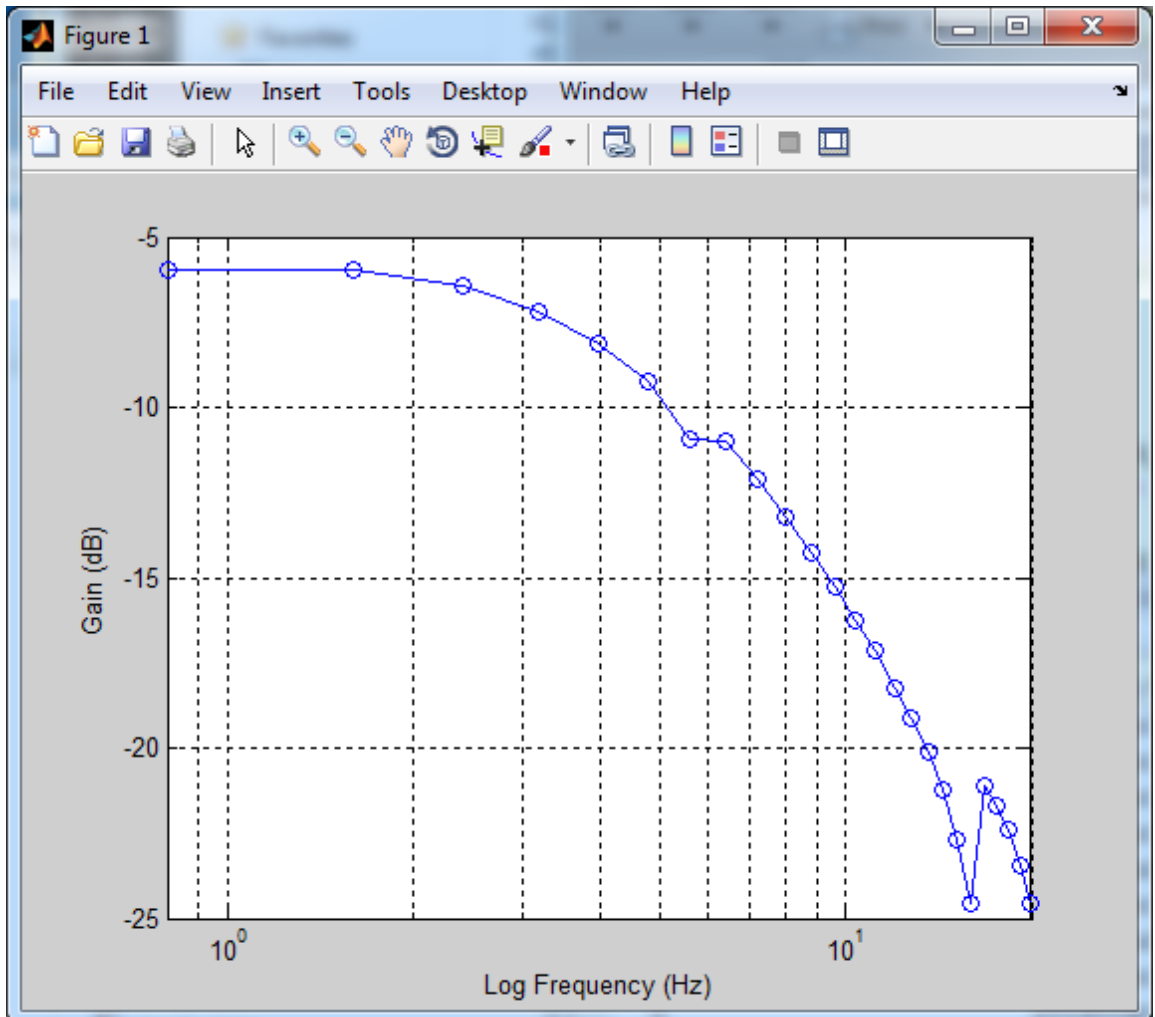

**Figure 8: Bode Plot (magnitude). The vertical axis is the ratio of output voltage (Vout, from the Baratron) to the input modulator input voltage (Vin) measured in Decibels. Decibels = 20* log10(Vout/Vin).**

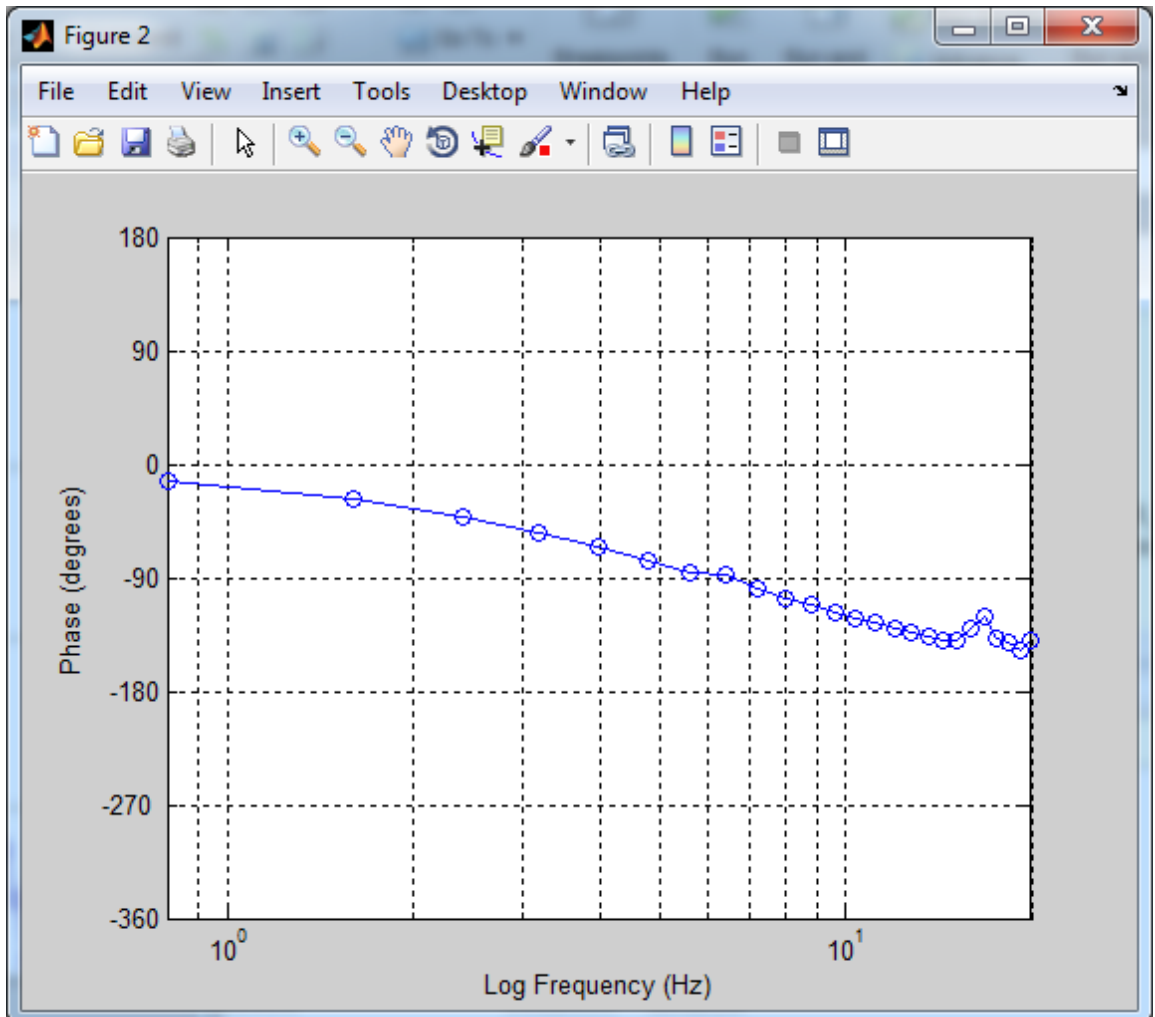
**Figure 9: Bode Plot (phase). The phase delay is given in degrees.**

## Fourier analysis of Turbine Pulse

The ultimate goal of this project was to generate an exact mirror of a real turbine pulse in a research chamber. Anything less would compromise future research having to do with human exposure thresholds and reactions to this infrasound while using our system as a research tool.

Figure 10 shows a wind turbine pulse extracted from previous work published at Wind Turbine Noise 2015 in Scotland.

In this case, we extracted a pulse by pointing an optical telescope at a turbine blade to get a tachometer signal and averaged together several observed pulses, separating them from the clutter of other turbines and wind noise as referenced earlier in this paper.

The turbine pulse we isolated has a period of 1.25 seconds or 0.8 cycles per second (Hertz). This pulse increases to a peak value, then decreases sharply to negative peak value, and then returns to zero.

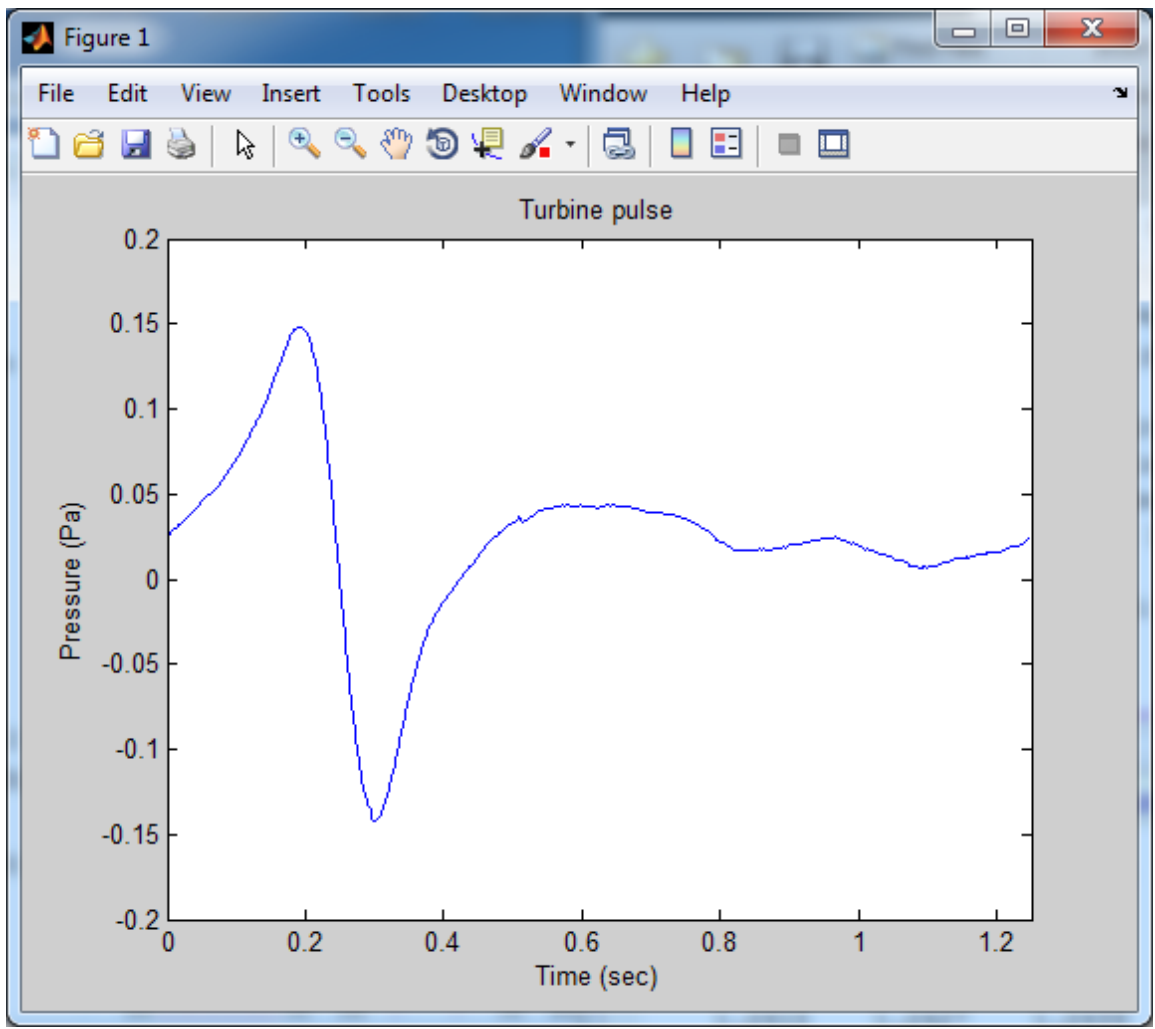

**Figure 10. A turbine pulse, from our paper (WTN 2015). The turbine pulse has a fundamental frequency of approximately 0.8Hz. The period is 1/f, about 1.25 seconds.**

This repeats over and over again in the signal. There is this general dipole nature. There are other properties of the signal which may also be observed, but this dipole nature is characteristic of the turbine signal.

The turbine signal is measured in Pascals so in this case, it goes from a positive value of 0.15 Pascals to negative value of 0.15 and has a zero value.

We can determine the level in decibels from the Pascal level using the formula:
Pressure in Decibels = 20* log10 (P/Pref)
where P=0.15 Pa, and Pref is the reference value 20x10^-6 Pa.
This gives a value of 77.5 dB (peak value) for this pulse.

We observed that the turbine pulses may be between 60 and 80 decibels (peak level) at standard setback distances.

We note that turbine pulses are a quasi periodic signal i.e.; the same signal repeats every period. The frequency is nearly constant, but it may speed up or slow down as the turbine changes speed. Furthermore, the exact shape of the turbine pulse may change, for example, as the turbine slowly changes direction with respect to our recording microphone. In this case, however, we are going to generate exactly the same turbine pulse over and over again. So it is a periodic signal with a fixed shape.

To understand our turbine pulse we appeal to Fourier analysis. Fourier analysis breaks a signal down into a series of sinusoids. If we have a periodic signal, we refer to these sinusoids as harmonics. So the first harmonic is at the frequency of the turbine signal so that it's at 0.8 Hz.

The second harmonic is at twice that 1.6 Hz. The third harmonic is at 2.4 Hz and so on.
In Figure 11 we show how to construct a turbine pulse from these sinusoidal components, these individual sine waves,

At the left are the individual sine waves so the top figure is a figure that has a frequency of 0.8 hertz and then below that is the second harmonic and the third harmonic all the way down to the seventh harmonic at the bottom. At the right is the summation of these components and we can see a turbine pulse. In this case there are two pulses. This is an interval from 0 to 2 seconds, so 2/1.25 or 1.6 periods are shown.

We can see a turbine pulse emerges as we add more and more components. And at the bottom of the figure, is shown the 25th component. That corresponds to 25 x 0.8Hz = 20 Hz and we see that the turbine pulse becomes extremely sharp at that frequency.

So the critical thing is to generate a turbine pulse we can determine how much of each harmonic needs to be added. And once we have that, we can construct a turbine pulse. Now how many harmonics are needed? Figure 12 shows the magnitude of each harmonic for the previous turbine pulse. We can see that the first harmonic - there's not very much energy in it. The energy is concentrated in the second and third harmonic and then falls off to zero. It appears that about a dozen harmonics are sufficient to capture that turbine shape.

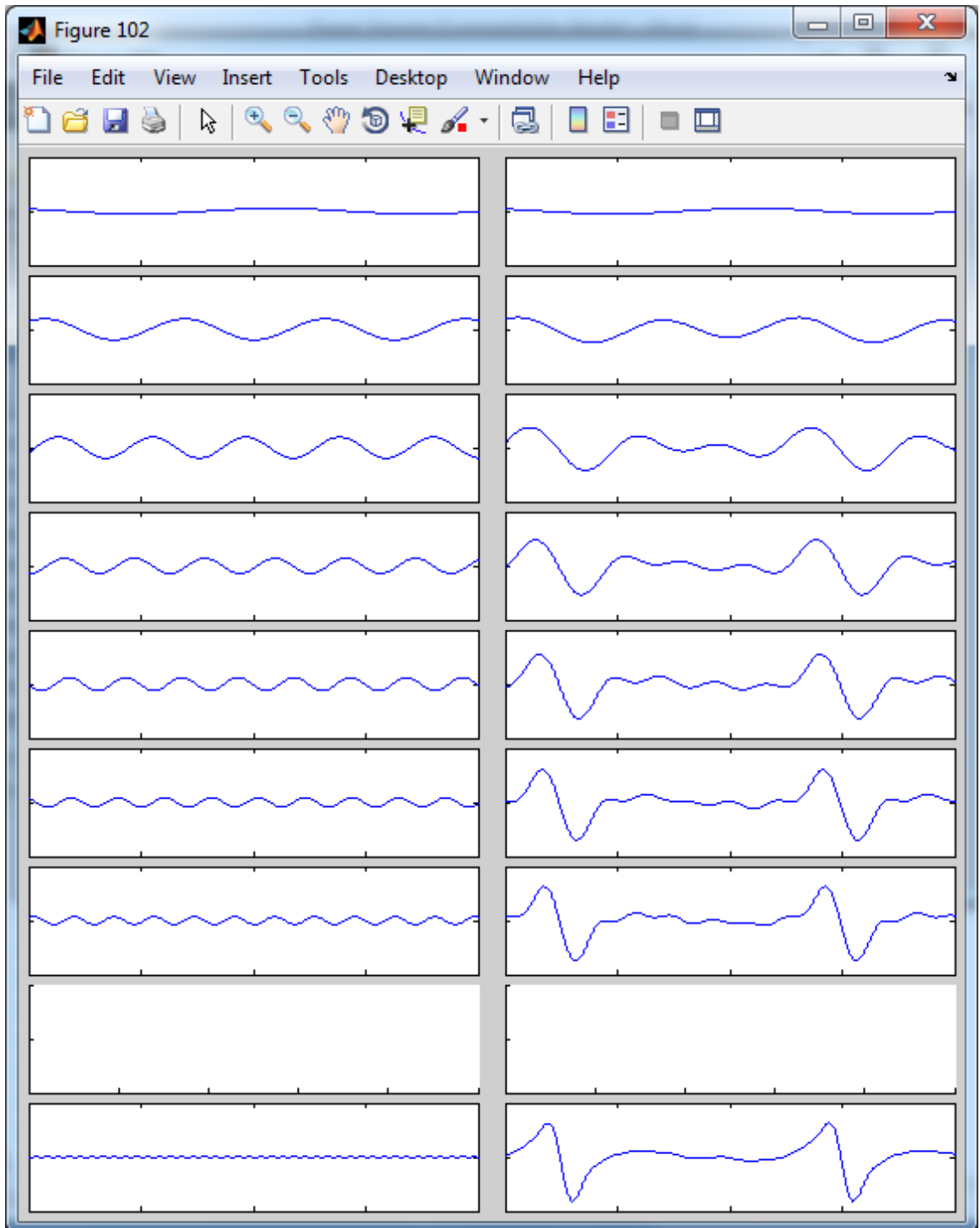

**Figure 11. Fourier analysis. Building up a turbine pulse by adding a number of sine waves. Sine waves are all integer multiples of the fundamental frequency, in this case f=0.8Hz. The left column shows each sine wave component from 0.8Hz, 1.6Hz, 2.4Hz, all the way to the seventh harmonic at 5.6Hz. The bottom row shows the 25th harmonic, which is at 20Hz. The right column shows the total wave. As harmonics are added the curve becomes sharper, and more like the turbine pulse.**

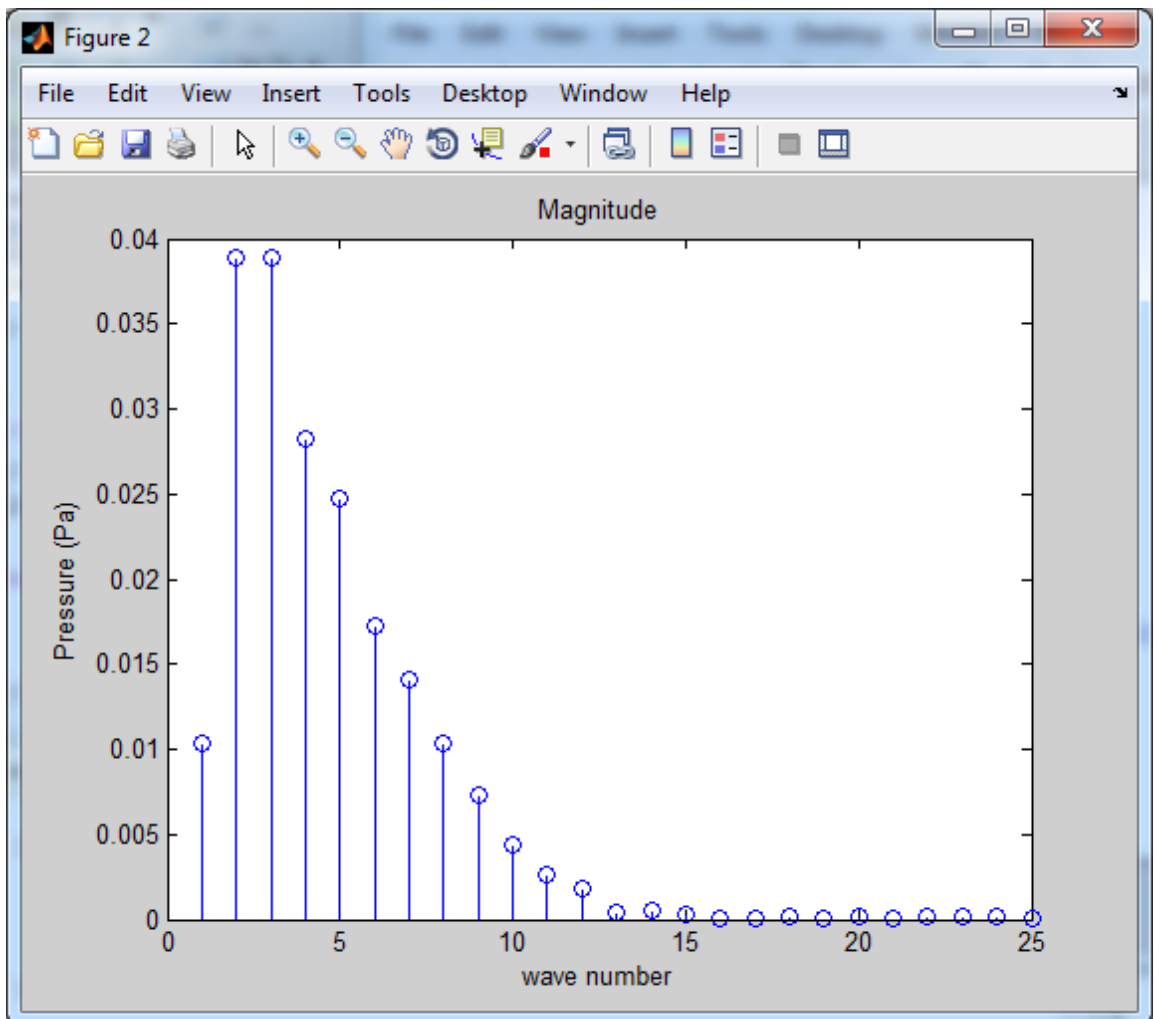

Figure 12. Fourier transform of the turbine pulse. Each bar is the magnitude of the kth harmonic, from 1 to 25. The vertical axis is sound pressure (Pascals). Note that most energy is contained in the second and third harmonics (1.6 Hz, 2.4Hz) and falls off with increasing frequency. A dozen harmonics are typically sufficient to capture this type of pulse.

# Generating an actual turbine Pulse in the Chamber

We showed that the frequency response of the chamber falls off as the frequency increases so it's not sufficient to just put these components alone into the chamber.

The higher frequency components won't be as strong in the chamber.

What we have to do, is look at our frequency response, and pre-compensate the signal before putting it into the chamber.

We are going to take the harmonics we had previously, increase the magnitude of the higher harmonics corresponding to the decrease we had in the frequency response, and thereby invert the effect of the frequency response.

Further, we are going to take the individual harmonics and shift them forward in time to reverse the delay. We are going to back up each of the harmonics to account for this delay.

Figure 12 shows what a pre-compensated signal looks like for this turbine pulse. The blue signal is the desired turbine pulse, and the red signal is the pre-compensated signal.

What you can see from this is that the red signal is accentuated at areas where it is varying quickly (high frequencies) and furthermore, the zero crossing of the red signal proceeds the zero crossing of the desired signal because we have to back up the signal to account for the delay of the chamber.

The idea is if we put this pre-compensated (red) signal in the chamber, we should get original turbine pulse out (the blue signal).

We generated 20 periodic waves (pre-compensated turbine pulses) and then we measured the chamber. The blue signal shows what we put in, and the red shows the signal recorded in the chamber.

If we zoom in on this, (Figure 14) we see, in fact, that the red signal corresponds to the turbine signal that we had measured verifying that we can generate this actual turbine signal in the chamber.

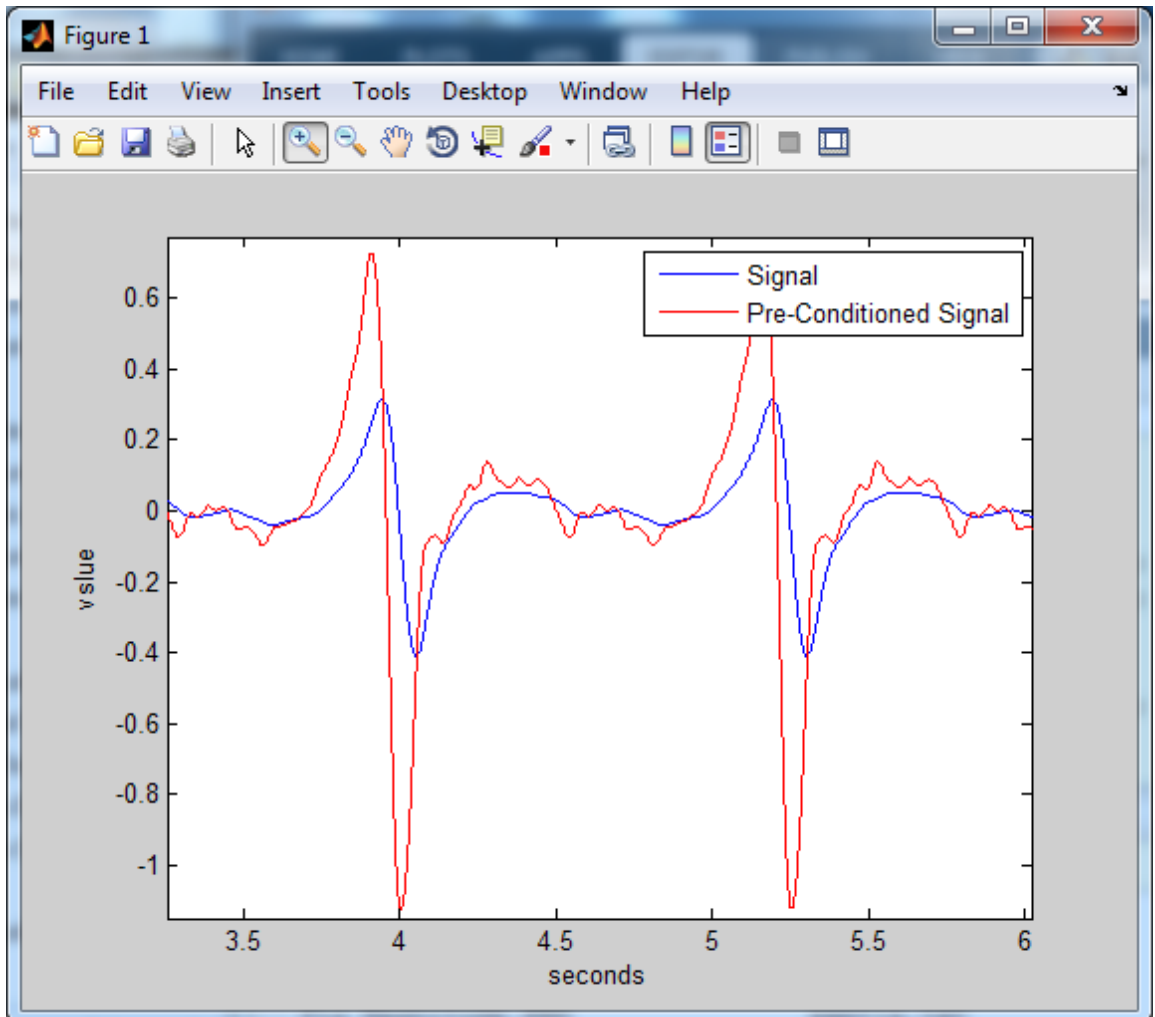

**Figure 12. Pre-compensated pulse.** To get the desired pulse in the chamber we need to **pre-compensate** each of the harmonic components. In particular, as harmonic k increases we need to increase the magnitude to compensate for the exact amount that response of the chamber decreases. This is visible as increased height of the steep changes on the turbine pulse. The higher frequency components are also advanced in time to correct for the (phase) delay of the chamber. Therefore the zero crossings precede the intended location.

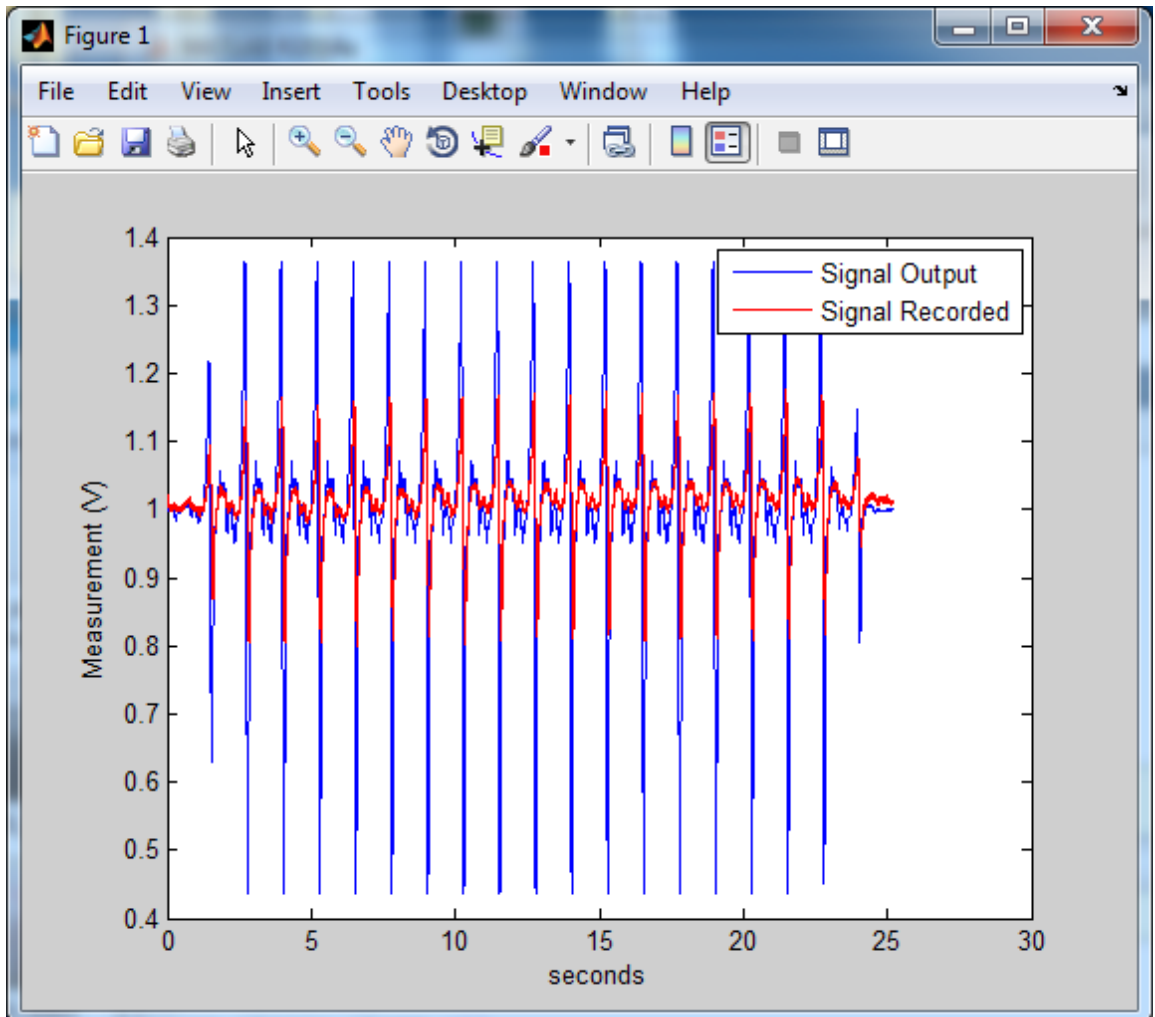

**Figure 13. Measured response. The chamber is driven with twenty (20) pre-compensated turbine pulses.**

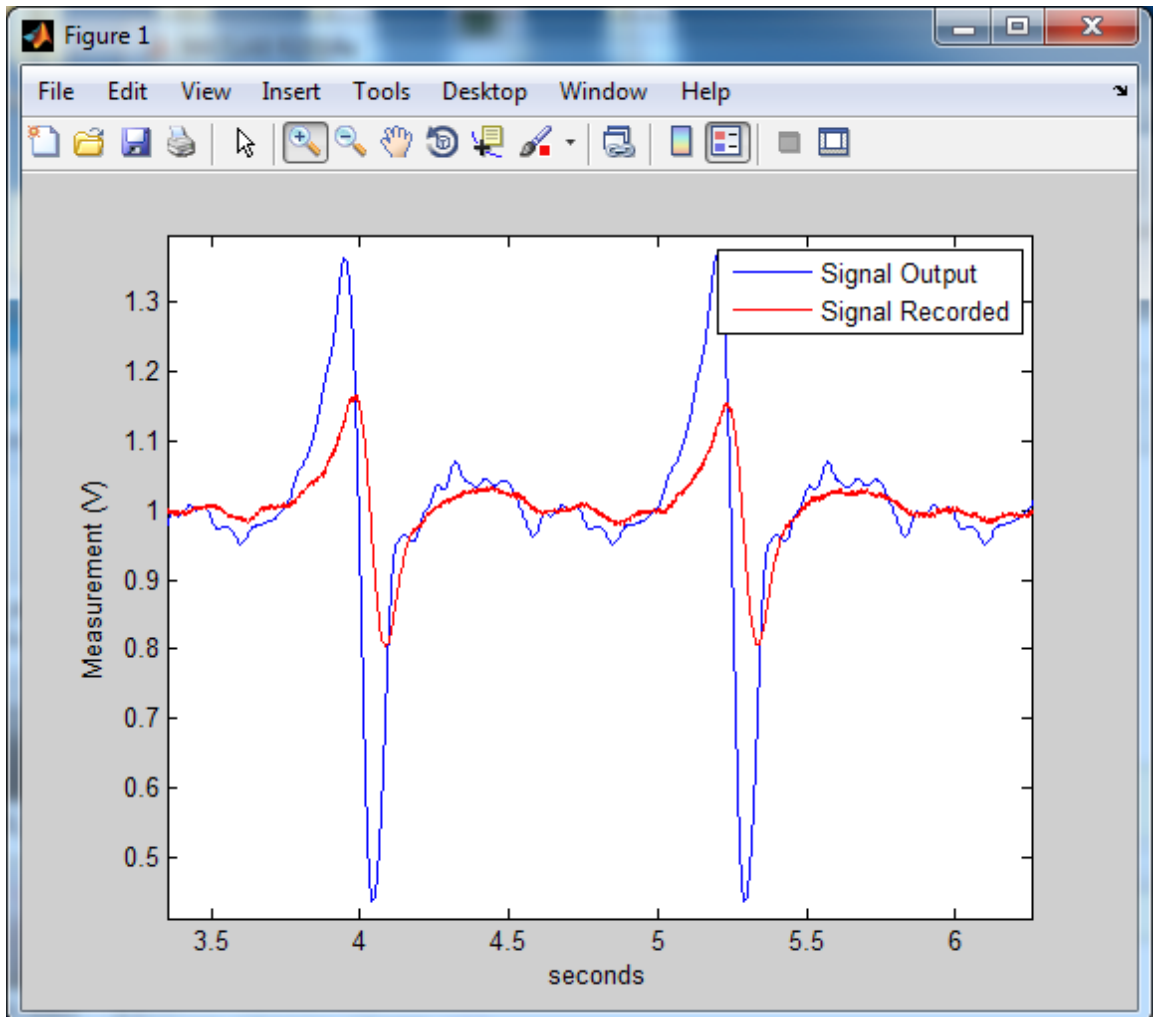

**Figure 14. Measured response. A blow up of the previous figure shows that the desired turbine pulse (red curve) is reconstructed in the chamber.**

Note: The level of this pulse is much greater than the turbine pulse measured.
To compute the level note that the measured signal (red curve) goes from approximately 0.8V to 1.15V. This corresponds to a peak amplitude of approx. 0.175V from the average value. The peak value in Pascals is then 0.175 V * 50Pa/V = 8.75 Pa. The value in dB is 20*log10 (8.75/20e-6) ~= 113 dB.

This is significantly higher than our measured turbine pulses at normal setback distances, which range from approximately 60 to 80dB.

Generated pulses at this level of course, would not be used for testing human thresholds and response.

Testing pulses would be restricted to the equivalent of turbine levels at approved setback distances as outlined in our previous description of chamber safety features.

It does prove however that our system is capable of replicating any turbine pulse that we come across and also potentially amplifying by a large factor.

## Appendix: Fourier Analysis

Fourier analysis. Any periodic signal $x(t)$ can be represented by a sum of sinusoids,
$$x(t)=\sum a_k \sin(2\pi f_0 kt + \phi_k)$$
where $k \geq 1$ is the wave number, $a_k$ is the magnitude and $\phi_k$ is the phase shift (radians) and $f_0 = 1/T$ is the fundamental frequency (0.8 Hz in this case).

The frequency response of the system (air source + chamber) is given by a set of magnitudes and phases, $b_k, \theta_k$ for each harmonic frequency $f_k = k f_0$.

The compensated signal $y(t)$ is then computed,
$$y(t)=\sum a_k/b_k \sin(2\pi f_k + \phi_k - \theta_k)$$

## Acknowledgements:

The authors would like to acknowledge the contribution of the following:

**George Dixon**
Vice President, Academic & Provost.  Previously Vice-President, University Research Professor. Department of Biology. University of Waterloo.
Who In October of 2015, while head of Office of Research, approved funding for this project.

**Mark Giesbrecht** Professor and Director, David R. Cheriton School of Computer Science, University of Waterloo.
For research support.

**Nicholas Kouwen**, P.Eng. Distinguished Professor Emeritus, Civil and Environmental Engineering, University of Waterloo
Who arranged much needed space in engineering for the chamber and related equipment so we could conduct further testing and system development at the university.

**Dr. John Vanderkooy** Distinguished Professor Emeritus Department of Physics and Astronomy University of Waterloo
For scientific advice, particularly with formalizing the mathematical underpinnings of sound and infrasound.

**Serhiy Yarusevych** Mechanical and Mechatronics Engineering, University of Waterloo.
For the loan of a Pitot Tube.

**Trevor Bonanno** of ELX Services, who entered into the spirit of the project after quoting a price for building the actual chamber. As a result of many revisions and extra requirements he willingly ended up finishing the final chamber at a loss instead of a profit.

**Bill Lassaline** of The Source Heating & Air Conditioning, who, realizing we were working on a very limited budget, sold us some material at his cost, made his "bone yard' of old furnaces available for parts, and even had his staff remove parts we needed from old furnaces.

**Mac Voisin & Bahar Marcela** Their generous donation allowed us to purchase the latest state of the art measurement equipment.